\documentclass[10pt,twocolumn,showpacs,showkeys,preprintnumbers,amssymb,aps,superscriptaddress]{revtex4-2}

\usepackage{amsmath, amssymb}
\usepackage{graphicx}
\usepackage{color,array}
\usepackage[colorlinks={true}]{hyperref}
\hypersetup{colorlinks=true,linkcolor=red,citecolor=magenta,urlcolor=blue}
\usepackage{comment}
\usepackage{xcolor}
\usepackage{color,array}
\usepackage{bm}
\usepackage{subfigure}
\usepackage{amsmath}
\usepackage{lipsum,appendix}
\usepackage{orcidlink}
\usepackage{pstricks}
\usepackage{booktabs}
\usepackage{amsmath}
\usepackage[margin=1in]{geometry}

\begin{document}

\title{
Experimental probe of quantum coherence in top-quark pair production { at the LHC}
}

\author{Saeed Haddadi~\!\!\orcidlink{0000-0002-1596-0763}}\email{haddadi@ipm.ir}
\address{School of Particles and Accelerators, Institute for Research in Fundamental Sciences (IPM), P.O. Box 19395-5531, Tehran, Iran}

\author{Majid Azizi~\!\!\orcidlink{0000-0002-1935-5923}}
\address{School of Particles and Accelerators, Institute for Research in Fundamental Sciences (IPM), P.O. Box 19395-5531, Tehran, Iran}

\author{Artur Czerwinski~\!\!\orcidlink{0000-0003-0625-8339}}
\affiliation{Institute of Physics, Faculty of Physics, Astronomy and Informatics, Nicolaus Copernicus University in Torun, ul. Grudziadzka 5, 87-100 Torun, Poland}
\affiliation{STARTOVA UMK Sp. z o.o., ul. Gagarina 7, 87-100 Torun, Poland}

\author{Hamid Arian Zad~\!\!\orcidlink{0000-0002-1348-1777}}  
\affiliation{Department of Theoretical Physics and Astrophysics, Faculty of Science of P. J. \v{S}af{\'a}rik University, Park Angelinum 9, 040 01 Ko\v{s}ice, Slovak Republic}


\begin{abstract}
{
We investigate quantum coherence in the spin degrees of freedom of top--antitop pairs produced at the LHC using the $l_1$-norm of coherence in the helicity basis. Combining Standard Model predictions with the spin density matrix reconstructed from the CMS measurement of top-quark spin correlation coefficients, we study the dependence of coherence on the invariant mass $M_{t\bar t}$ and production angle. To quantify the experimental precision of the extracted coherence, we propagate the measured uncertainties and covariance matrix of the spin-correlation coefficients through a Monte Carlo procedure. We find a non-monotonic kinematic dependence: coherence is large near the production threshold, is reduced in the intermediate-mass region, and increases again in boosted central events. The theoretical predictions are in good agreement with the CMS determinations in all three kinematic regions. Moreover, the experimentally reconstructed coherence is nonzero with a significance exceeding $5\sigma$ in each region. In particular, in the intermediate-mass region, $400<M_{t\bar t}<600~\mathrm{GeV}$, we obtain $C_{l_1}=0.597\pm0.037$. This result is especially noteworthy because previous analyses of the same reconstructed top-quark spin density matrix found concurrence and negativity, both entanglement measures, to be consistent with zero, $0.00\pm0.02$, in this kinematic region. The persistence of sizable coherence where entanglement is not experimentally certified demonstrates that coherence provides complementary information on the off-diagonal helicity-interference structure of the top-quark spin density matrix. Our results establish quantum coherence as an experimentally accessible and statistically robust observable for probing the quantum structure of top-quark pair production at the LHC and provide a basis for future precision tests of quantum chromodynamics and searches for anomalous top-quark interactions.
}
\end{abstract}

\maketitle

\section{Introduction} \label{sec:Introduction}
As the heaviest known elementary particle, the top quark occupies a unique position within the Standard Model (SM). Because it decays before hadronization occurs, its spin information is directly encoded in the kinematic distributions of its decay products, providing a unique opportunity to study spin dynamics and quantum correlations in a fundamentally relativistic regime. Consequently, top--antitop ($t\bar t$) production at the LHC provides a powerful framework for investigating quantum correlations, interference effects, and spin dynamics in high-energy collisions. Measurements of top-quark spin correlations now constitute an important component of the LHC physics program, offering precision tests of the SM and sensitivity to higher-order quantum chromodynamics (QCD) effects as well as possible contributions from non-standard interactions.~\cite{Afik2021,Afik2022quantuminformation,CMSPRD2024}.

In recent years, concepts from quantum information theory have been successfully incorporated into collider phenomenology. Since the spin state of a $t\bar t$ pair can be described as a two-qubit quantum system, tools originally developed in quantum information have found natural applications in high-energy physics. This perspective has led to extensive investigations of entanglement, Bell nonlocality, quantum steering, quantum discord, and quantum tomography in top-quark pair production and related particle systems~\cite{Afik2021,Afik2022quantuminformation,PhysRevLett.130.221801,Severi2022,Aguilar-Saavedra2022,PhysRevLett.127.161801,PhysRevD.109.115023,Han2024}. In particular, the proposal of quantum-state tomography for top-quark pairs~\cite{Afik2021} and the subsequent experimental reconstruction of the spin-density matrix by the CMS Collaboration~\cite{CMSPRD2024} have established a direct connection between collider observables and the language of quantum information.

Among the various quantum resources that can be encoded in a density matrix, quantum coherence occupies a particularly fundamental role. Quantum coherence originates from the existence of off-diagonal density matrix elements and directly quantifies the quantum superpositions responsible for interference phenomena~\cite{PhysRevLett.113.140401,RevModPhys.89.041003}. Within the resource-theoretic framework of quantum information, coherence is regarded as a more general property than entanglement, since entanglement requires coherence in an appropriate basis whereas coherent states need not be entangled~\cite{PhysRevLett.113.140401,RevModPhys.89.041003}. More generally, the hierarchy of quantum resources places quantum coherence as the most fundamental and general manifestation of quantumness, encompassing quantum discord, entanglement, quantum steering, and Bell nonlocality as progressively stronger forms of quantum correlations~\cite{PhysRevLett.98.140402,RevModPhys.81.865,Adesso_2016,HU20181}. As a result, coherence can remain substantial even when standard entanglement criteria fail, providing access to aspects of the quantum state that are invisible to entanglement measures alone. { An important consequence of this distinction is that quantum coherence may remain experimentally observable even in kinematic regions where entanglement witnesses no longer certify non-separability. Since the two observables probe different structural properties of the reconstructed density matrix, coherence can reveal residual helicity interference that is invisible to entanglement measures. Establishing whether such complementary information can be accessed experimentally is therefore an important motivation for introducing coherence as a collider observable.}

The study of quantum correlations in high-energy physics has advanced rapidly in recent years. Entanglement has been predicted and analyzed in top-quark pair production~\cite{Afik2021,Afik2022quantuminformation,Severi2022,Severi2023}, experimentally observed by the ATLAS and CMS Collaborations~\cite{ATLAS2024,CMSCollaboration_2024,CMSPRD2024}, and proposed as a sensitive probe of new interactions and effective-field-theory effects~\cite{Fabbrichesi2023,PhysRevD.106.055007,JHEO2024}. Bell nonlocality and related quantum information observables have likewise been investigated in top-quark systems~\cite{PhysRevLett.127.161801,Aguilar-Saavedra2022,Han2024,PhysRevD.111.033004} as well as in other particle processes involving mesons, hyperons, quarks, and charmonia~\cite{PhysRevD.109.L031104,PhysRevD.110.053008,Cheng:2025cuv,PhysRevD.110.054012}. Nevertheless, most existing studies focus on determining whether a given quantum state satisfies specific entanglement or nonlocality criteria \cite{Afik2022quantuminformation}. By contrast, the behavior of quantum coherence that is a quantity directly associated with the interference structure of the production amplitudes, has received less attention, despite its more general character and its direct connection to the off-diagonal elements of the density matrix.

The top-quark sector provides a particularly attractive environment for investigating quantum coherence. The spin density matrix describing the produced $t\bar t$ ensemble is determined by the helicity amplitudes of the underlying QCD processes, predominantly gluon--gluon fusion and, to a lesser extent, quark--antiquark annihilation. The off-diagonal elements of this density matrix arise from interference among different helicity configurations and therefore encode information about the Lorentz and gauge structure of the top-quark production mechanism. Quantum coherence thus offers a direct probe of the helicity-interference pattern predicted by QCD and provides a complementary perspective on the spin structure of top-quark pair production. Recent theoretical studies have begun exploring coherence-related observables in top-quark systems~\cite{coherence2025,CHEN2026140426}, further motivating a systematic investigation within experimentally accessible collider observables.

{
In this work, we investigate the quantum coherence encoded in the spin degrees of freedom of top--antitop pairs produced at the LHC. Using the $l_1$-norm of coherence~\cite{PhysRevLett.113.140401} as a quantitative measure, we analyze its dependence on the invariant mass of the $t\bar t$ system and on the production angle in the helicity basis. Our study is motivated by the recent CMS measurement of the complete set of spin-correlation coefficients~\cite{CMSPRD2024}, which enables the reconstruction of the spin density matrix and, consequently, the extraction of experimentally observable coherence measures. To quantify the experimental uncertainties of the extracted coherence, we employ a Monte Carlo uncertainty-propagation procedure based on the measured uncertainties and covariance matrix of the spin-correlation coefficients. This allows us to perform a direct comparison between SM predictions and experimental data. We find that quantum coherence exhibits a nontrivial kinematic dependence. It is large near the production threshold, partially suppressed in the intermediate-mass region, and subsequently restored in boosted central events at high invariant mass. This behavior reflects the underlying helicity-interference structure of the QCD production amplitudes rather than a simple monotonic suppression due to increasing phase space. Furthermore, we show that significant coherence persists in regions where quantum discord, entanglement and so on are known to become weak or undetectable~\cite{fabbrichesi2025localvsnonlocalentanglement}, demonstrating that coherence captures additional information about the spin density matrix beyond conventional quantum correlation observables. The strong agreement observed between theoretical predictions and CMS measurements establishes quantum coherence as an experimentally accessible observable in top-quark physics. More broadly, our results demonstrate that coherence provides a complementary characterization of the quantum structure of $t\bar t$ production, linking collider observables directly to the interference properties of QCD amplitudes and extending the growing program of quantum information studies at high-energy colliders~\cite{Afik2022quantuminformation,BARR2024104134,Georgescu2024}.}

The remainder of this paper is organized as follows. In Sec.~\ref{sec:Methods}, we present the theoretical framework for describing the spin density matrix of top--antitop pairs and define the quantum coherence measure employed in our analysis. Section~\ref{sec:results} contains the numerical results, the comparison with CMS measurements, and their physical interpretation. Finally, concluding remarks and perspectives are presented in Sec.~\ref{sec:summary}.

\section{Method} \label{sec:Methods}

Our analysis is based on the density operator description of $t\bar t$ production at the LHC. Since the top quark decays before hadronization, the spin state of the produced $t\bar t$ system can be reconstructed from the angular distributions of its decay products. This feature makes top-quark pair production a particularly suitable framework for applying quantum information concepts to collider observables.

The spin degrees of freedom of the top and antitop quarks form an effective two-qubit system. The corresponding density matrix can be expressed in the Hilbert--Schmidt basis as
\begin{align}
	\label{Eq:density matrix}
	\rho = \frac{1}{4} \bigg(
	\mathbb{I}_2 \otimes \mathbb{I}_2
	&+ \sum_{i=1}^{3} B_i^{+}\,\sigma_i \otimes \mathbb{I}_2
	+ \sum_{j=1}^{3} B_j^{-}\,\mathbb{I}_2 \otimes \sigma_j
	\nonumber\\
	&+\sum_{i,j=1}^{3} C_{ij}\,\sigma_i \otimes \sigma_j
	\bigg),
\end{align}
where $\sigma_i$ denote the Pauli matrices and $\mathbb{I}_2$ is the $2\times2$ identity matrix. The vectors $B_i^{+}$ and $B_i^{-}$ describe the polarizations of the top quark and antiquark, respectively, while the coefficients $C_{ij}$ form the correlation matrix. These quantities completely characterize the quantum state of the $t\bar t$ system. The elements of $C_{ij}$ encode the correlations and quantum interference effects that are experimentally reconstructed from angular correlations among the top-quark decay products.

The density matrix in Eq.~(\ref{Eq:density matrix}) is fully specified by the 15 real parameters $B_i^\pm$ and $C_{ij}$. In the language of quantum information, these quantities correspond to expectation values of the observables $\sigma_i\otimes\mathbb{I}_2$, $\mathbb{I}_2\otimes\sigma_i$, and $\sigma_i\otimes\sigma_j$. Consequently, a complete determination of all coefficients allows the full reconstruction of the two-qubit quantum state, a procedure known as quantum state tomography~\cite{Afik2021}. In the context of top--antitop production, the required coefficients can be extracted from measured angular correlations in the decay products of the top and antitop quarks. Recent CMS measurements of the complete set of spin-polarization and correlation coefficients~\cite{CMSPRD2024} therefore enable the reconstruction of the density matrix (\ref{Eq:density matrix})  and provide direct experimental access to  the quantum observables derived from it. The coherence measure investigated in this work is obtained from this reconstructed density matrix and may thus be regarded as a tomographic observable that probes the interference structure of the underlying production amplitudes. { The experimental procedure consists of two conceptually distinct steps. First, the complete spin density matrix is reconstructed from the measured spin-polarization and spin correlation coefficients using quantum state tomography. Second, the $l_1$-norm of coherence is evaluated from the reconstructed density matrix. Thus, quantum coherence is not measured directly; rather, it is a derived quantum information observable extracted from experimentally reconstructed quantum states.}

In the partonic level at the LHC, top--antitop pairs are produced predominantly through gluon--gluon fusion,
\[
gg \rightarrow t\bar t,
\]
with a smaller contribution arising from quark--antiquark annihilation,
\[
q\bar q \rightarrow t\bar t.
\]
For proton--proton collisions at the LHC ($\sqrt{s}=13~\mathrm{TeV}$), approximately $90\%$ of the total production rate originates from the gluon-fusion channel, while the remaining contribution arises from quark--antiquark interactions~\cite{PhysRevD.110.030001}. Accordingly, in constructing the final spin-density matrix, we take into account the contributions from both production channels, with a weight of $90\%$ for the gluon-fusion process and $10\%$ for the quark--antiquark process.  The dominance of gluon fusion plays an important role in determining the spin structure and quantum correlations of the produced $t\bar t$ ensemble. 
\\
In the $t\bar t$ center-of-mass frame, the production kinematics are characterized by the invariant mass $M_{t\bar t}$ and the scattering angle $\Theta$. A convenient parameter describing the kinematics is the top-quark velocity
\begin{equation}
	\beta=\sqrt{1-\frac{4m_t^2}{M_{t\bar t}^{\,2}}},
\end{equation}
where $m_t$ denotes the top-quark mass. The production threshold corresponds to $\beta=0$, or equivalently
\begin{equation}
	M_{t\bar t}=2m_t\simeq346~\mathrm{GeV}.
\end{equation}
At leading order in QCD, the net polarizations vanish, and the spin density matrix is therefore fully specified by five independent spin-correlation coefficients. The analytical expressions for these coefficients in both the $gg$ and $q\bar q$ production channels are summarized in Appendix~\ref{appendixA}. These coefficients determine the density matrix in Eq.~(\ref{Eq:density matrix}) and therefore completely specify the quantum state of the produced top--antitop pair.

To quantify the quantum coherence encoded in the reconstructed spin-density matrix, we employ the $l_1$-norm of coherence, one of the standard measures in the resource theory of quantum coherence~\cite{PhysRevLett.113.140401}. For a density matrix $\rho$, it is defined as
\begin{equation}
	\mathcal{C}_{l_1}(\rho)	=\sum_{i\neq j}|\rho_{ij}|,
\end{equation}
where the summation extends over all off-diagonal matrix elements of $\rho$ in a fixed reference basis. 

{ It is important to emphasize that the $l_1$-norm of coherence is explicitly basis dependent, since it is defined through the off-diagonal elements of the density matrix in a chosen reference basis. Consequently, different spin bases generally yield different numerical values of coherence while describing the same underlying quantum state. This basis dependence reflects the fact that quantum coherence quantifies superposition relative to a specified set of basis states rather than representing an invariant property of the density matrix. Throughout this work, we employ the helicity basis because it provides the most natural description of top-quark pair production at the LHC. In this basis, the density matrix is directly determined by the helicity amplitudes of the underlying QCD processes, and the off-diagonal matrix elements quantify the interference between different helicity configurations. The resulting coherence therefore directly probes the helicity-interference structure of the production amplitudes, giving the observable a clear physical interpretation.}

\begin{figure}[t]
	\begin{center}
		\includegraphics[width=0.48\textwidth]{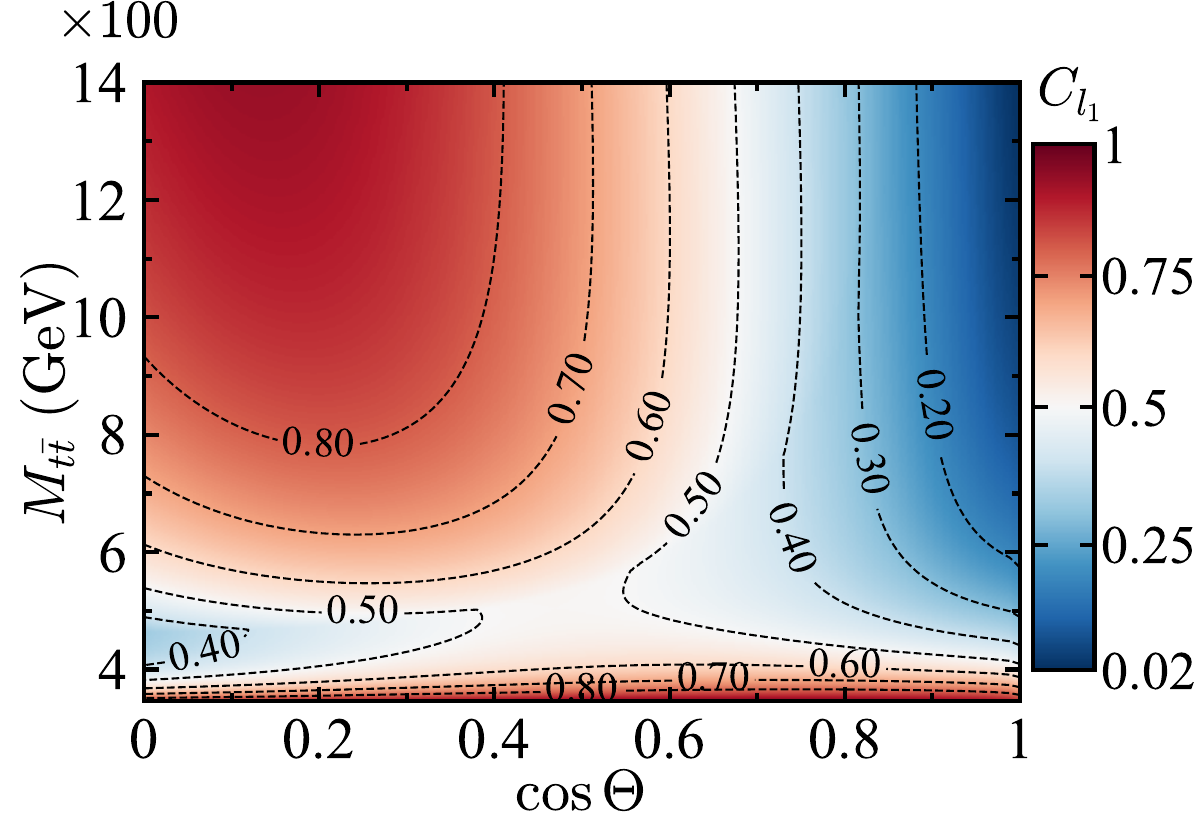}
	\end{center}
    \vspace{-0.4cm}
	\caption{The $l_1$-norm of quantum coherence, $C_{l_1}$, for top--antitop ($t\bar t$) pairs produced at the LHC as a function of the invariant mass $M_{t\bar t}$ and the production angle $\Theta$ { in the helicity basis}. 
    Large coherence is observed both near the production threshold and in the boosted central region, illustrating the nontrivial kinematic dependence of the spin-interference structure in top-quark pair production.}
	\label{figure1}
\end{figure}

\section{Results and discussion}
\label{sec:results}
Figure~\ref{figure1} provides the dependence of the $l_1$-norm of quantum coherence on the scattering angle (through $\cos\Theta$) and on the invariant mass  $M_{t\bar{t}}$ of the top–antitop quark pair produced at the LHC. In the low-mass regime ($346<M_{t\bar{t}}<400$ GeV), the $t\bar{t}$ pair is produced close to threshold, where the available phase space is limited and spin correlations between the quark and antiquark remain strong. The $l_1$-norm of coherence is relatively high ($C_{l_1}=0.754\pm0.111$ in Table~\ref{Table:1}), with small angular dispersion. This high coherence near threshold can be attributed to the near-resonant production channel, where the system remains dynamically correlated and less affected by random phase fluctuations from the partonic environment. As the invariant mass increases (intermediate regime, $400<M_{t\bar{t}}<600$ GeV), the phase space for gluon radiation and higher-order QCD effects becomes larger, leading to a partial loss of quantum coherence ($0.581\pm0.119$). The average $l_1$-norm of coherence decreases moderately, and the dependence on $\cos\Theta$ becomes more pronounced. This suggests that forward- and backward-scattering events begin to differ significantly in their coherent character.  Although coherence decreases, it remains nonzero across all scattering angles, indicating the persistence of intrinsic quantum superpositions within the production channel. Next, events with central scattering angles $(\cos\Theta<0.4$) and large invariant masses ($M_{t\bar t}>800\ \mathrm{GeV}$) show consistently large values of the $l_{1}$-norm of quantum coherence ($0.939\pm0.045$). The coherence in this boosted central bin has a small dispersion, indicating a robust, repeatable effect in that kinematic window. One might expect that raising $M_{t\bar t}$ increases radiation and therefore decoherence. Instead, the calculations show that for central angles, increasing the invariant mass does not destroy coherence, but rather enhances or preserves it. The small spread (low variance) of coherence in this region shows that the constructive interference is not a rare fluctuation but a stable feature of the underlying production dynamics at high $M_{t\bar t}$.

\begin{table}[t]
\centering
\caption{{ The $l_1$-norm of quantum coherence, $C_{l_1}$, for top--antitop pairs in the three kinematic regions considered in this work. The theoretical values are obtained from the SM spin-density matrix and correspond to the mean values over the respective regions shown in Fig.~\ref{figure1}. The experimental values are reconstructed from the CMS measurements of the spin-correlation coefficients. The quoted uncertainties are obtained by propagating the experimental uncertainties of the measured $C_{ij}$ coefficients, including their reported covariance matrix.}}
\vspace{0.5em}
\resizebox{\columnwidth}{!}{
\begin{tabular}{@{} l l c c c @{}}
\toprule
& & & \multicolumn{2}{c}{$\boldsymbol{C_{l_1}}$} \\
\cmidrule(lr){4-5}
Bins & \textbf{$M_{t\bar{t}}$ (GeV)} & \textbf{$\cos\Theta$} 
& \quad\textbf{Theor.}\quad & \textbf{Exp.} \\
\midrule
\hline
Threshold & $346-400$   & $0-1$  & $0.754\pm0.111$ & $0.803\pm0.098$ \\
Intermediate & $400-600$ & $0-1$ & $0.581\pm0.119$ & $0.597\pm0.037$ \\
Boosted central & $>800$ & $<0.4$ & $0.939\pm0.045$ & $1.006\pm0.113$ \\
\bottomrule
\end{tabular}}
\label{Table:1}
\end{table}

The complete set of coefficients $B_i^{\pm}$ and $C_{ij}$ appearing in Eq.~\eqref{Eq:density matrix}, which fully characterize the spin density matrix of the top–antitop ensemble, has recently been reported by the CMS Collaboration~\cite{CMSPRD2024}. This enables a direct comparison between our analytic predictions from the previous section and experimental data~\footnote{We use the observables  $C_{nr}^{\pm} = C_{nr} \pm C_{rn}$,  $C_{rk}^{\pm} = C_{rk} \pm C_{kr}$, and  $C_{nk}^{\pm} = C_{nk} \pm C_{kn}$, where the associated covariances are fully taken into account in the calculation~\cite{CMSPRD2024}.
}.  To perform this test, we recast the CMS measurements in terms of the corresponding quantum coherence values obtained on the helicity basis, as summarized in Table~\ref{Table:1}. 
{ To estimate the experimental uncertainty of the $l_1$-norm of quantum coherence, we employ a Monte Carlo uncertainty-propagation procedure based on the covariance matrices of the measured spin correlation coefficients $C_{ij}$ provided by the CMS Collaboration through the associated HEPData record~\footnote{\href{https://hepdata.net/record/ins2829523}{HEPdata, https://hepdata.net/record/ins2829523}}. The correlations among the $C_{ij}$ coefficients are fully taken into account in the sampling procedure. Correlations involving the polarization coefficients $B_i^{\pm}$ and the normalization coefficient $c$, as well as correlations between observables belonging to kinematic bins not considered in the present analysis, are neglected.}
This table shows an overall strong consistency between the theoretical predictions (Fig.~\ref{figure1}) and the CMS experimental measurements of the $l_{1}$-norm of quantum coherence, with the level of agreement depending on the kinematic regime.

{
In the near-threshold region ($346<M_{t\bar t}<400$~GeV), the theoretical prediction for the $l_1$-norm of quantum coherence is $C_{l_1}^{\rm th}=0.754\pm0.111$, while the value reconstructed from the CMS data is $C_{l_1}^{\rm exp}=0.803\pm0.098$. The central values differ by approximately $6.5\%$ relative to the theoretical prediction, corresponding to a difference of only $0.049$.  Taking the quoted uncertainties into account, the difference between the theoretical and experimental central values is only approximately $0.33\delta_{\rm comb}$, where the combined uncertainty is obtained as $\delta_{\rm comb}=\sqrt{\delta_{\rm th}^2+\delta_{\rm exp}^2}$.
This excellent agreement demonstrates that the theoretical framework provides a reliable description of the quantum coherence in the near-threshold region. In particular, the relatively large value of $C_{l_1}$ indicates strong helicity interference and spin correlations close to the production threshold, where the available phase space is restricted and the effects of additional QCD radiation are comparatively limited. The consistency between the theoretical prediction and the experimentally reconstructed coherence therefore provides quantitative support for the underlying QCD description of the spin-density matrix in this kinematic regime.}

{
In the intermediate-mass region ($400<M_{t\bar t}<600$~GeV), the theoretical prediction for the $l_1$-norm of quantum coherence is $C_{l_1}^{\rm th}=0.581\pm0.119$, while the value reconstructed from the CMS data is $C_{l_1}^{\rm exp}=0.597\pm0.037$. The two central values differ by only $0.016$, corresponding to a relative deviation of approximately $2.8\%$ with respect to the theoretical prediction. Taking the quoted uncertainties into account, the difference between the theoretical and experimental central values is only approximately $0.13\delta_{\rm comb}$. Thus, the theoretical prediction and the experimental determination are in excellent agreement also in this kinematic region. The relatively reduced coherence compared with the threshold region is consistent with an enhanced sensitivity to higher-order QCD effects, gluon radiation, and angular averaging, which can modify the interference structure encoded in the spin-density matrix.
Importantly, this intermediate-mass region coincides with the kinematic domain in which previous studies based on the reconstructed top-quark spin density matrix found that conventional entanglement measures cease to detect entanglement~\cite{Afik2022quantuminformation,fabbrichesi2025localvsnonlocalentanglement}. In Ref.~\cite{fabbrichesi2025localvsnonlocalentanglement}, the reconstructed spin density matrix in this region yields two widely used entanglement measures, namely concurrence and negativity, both consistent with zero ($0.00\pm0.02$), indicating that quantum entanglement cannot be experimentally certified within the quoted uncertainties. In contrast, quantum discord, which quantifies more general quantum correlations beyond entanglement, remains finite with a value of $0.097\pm0.013$. Remarkably, the experimentally reconstructed $l_1$-norm of quantum coherence also remains substantial, with $C_{l_1}=0.597\pm0.037$. Thus, while concurrence and negativity are consistent with zero, both quantum discord and quantum coherence remain finite in the same kinematic region.
This comparison demonstrates that the absence of detectable entanglement does not imply the absence of quantum superposition or interference in the reconstructed state. Rather, the sizable value of $C_{l_1}$ indicates that the off-diagonal elements of the spin density matrix continue to encode significant helicity interference generated by the QCD production amplitudes. In this sense, quantum coherence provides complementary information beyond conventional entanglement measures and remains experimentally accessible in a kinematic region where entanglement-based observables provide limited discrimination. The persistence of a sizable coherence in this regime therefore extends the experimentally accessible quantum information characterization of top-quark pair production and highlights the potential of coherence as a complementary probe of the underlying spin-interference structure.}

{ 
In the boosted-central region, characterized by $M_{t\bar t}>800$~GeV and $\cos\Theta<0.4$, the theoretical and experimental values of the $l_1$-norm of quantum coherence are $C_{l_1}^{\rm th}=0.939\pm0.045$ and $C_{l_1}^{\rm exp}=1.006\pm0.113$, respectively. The central values differ by approximately $7.1\%$, while the difference corresponds to only about $0.55\delta_{\rm comb}$ when the quoted uncertainties are combined in quadrature, demonstrating good consistency between the theoretical prediction and the CMS determination. The large value of $C_{l_1}$ in this region further shows that quantum coherence is not necessarily suppressed at large invariant mass; instead, it can remain substantial in specific angular configurations. This behavior reflects the persistence of strong helicity interference in boosted central $t\bar t$ production and demonstrates that the underlying spin-interference structure is well described by the SM.
}

{
The propagated experimental uncertainties further allow us to quantify the significance of the nonzero quantum coherence in each kinematic region. Following the standard convention commonly used in collider analyses, we estimate the significance from the ratio of the experimentally reconstructed central value to its propagated standard uncertainty.
This yields  $8.2\sigma$, $16.1\sigma$, and $8.9\sigma$ for the threshold, intermediate-mass, and boosted-central regions, respectively. Thus, the experimentally reconstructed quantum coherence is nonzero with a significance exceeding $5\sigma$ in all three kinematic regions. In particular, the intermediate-mass region exhibits the highest significance, owing to its relatively small propagated experimental uncertainty. These results demonstrate that the nonzero coherence extracted from the CMS spin-density matrix is statistically robust across all three kinematic regimes.
}

The analysis of quantum coherence in top--antitop production offers a direct and highly nontrivial probe of the spin dynamics predicted by the SM. In proton--proton collisions at the LHC, $t\bar t$ pairs are produced predominantly via gluon--gluon fusion, a process dictated by the SU(3)$_c$ gauge symmetry of QCD~\footnote{SU(3)$_c$ gauge symmetry is the non-Abelian color symmetry underlying QCD, whose vector top--gluon interaction uniquely fixes the helicity amplitudes and thus determines the spin-interference structure of top--antitop production}. The Lorentz and gauge structure of the top--gluon vertex fully determines the helicity amplitudes and, consequently, the form of the spin density matrix describing the produced $t\bar t$ system. The off-diagonal components of this density matrix, quantified through the $l_1$-norm of coherence, originate from interference among distinct helicity configurations in the QCD production amplitude. Hence, both the magnitude and the kinematic behavior of quantum coherence provide a direct window into the vector character of the gluon coupling, angular momentum conservation, and the unitarity of the underlying quantum field-theoretic framework.

Because the $l_1$-norm of coherence directly probes the off-diagonal elements of the spin-density matrix, it is particularly sensitive to interactions that modify the helicity structure of the production amplitudes. In an effective-field-theory description, examples include the chromomagnetic and chromoelectric dipole operators, which alter the interference pattern among top-quark helicity states and can therefore induce characteristic distortions in the coherence distribution as a function of $M_{t\bar t}$ and $\cos\Theta$. The kinematic regions identified in this work, especially the boosted-central regime where coherence is maximized and the intermediate-mass window where it is most sensitive to radiative effects, provide natural benchmarks for future studies of such effects.

In this study, quantum coherence serves as a complementary observable to traditional spin-correlation analyses. While entanglement measurements probe the nonseparability of the state, coherence quantifies the interference structure encoded in the production amplitudes. The strong agreement between theoretical predictions and CMS data reported in Table~\ref{Table:1} provides further independent support for the SM description of top-quark pair production at TeV-scale energies. The observation that coherence remains robust in both the near-threshold and boosted-central regimes is consistent with the SM expectation that spin correlations arise from coherent quantum superpositions of partonic helicity states rather than from incoherent classical mixtures.

Some studies~\cite{PhysRevD.106.055007,Severi2023,JHEO2024,CHEN2026140426} have demonstrated that quantum correlations provide a powerful and sensitive probe of physics beyond the SM. The results presented here further support this direction, as quantum coherence represents a more general quantum resource that encompasses a broader class of nonclassical features within the hierarchy of quantum correlations. Consequently, deviations in the dependence of coherence on $M_{t\bar t}$ or $\cos\Theta$ could provide complementary signatures of new physics, including anomalous chromomagnetic or chromoelectric dipole couplings and contributions from higher-dimensional operators within an effective-field-theory description.


\section{Summary}
\label{sec:summary}
{ In this work, we have investigated quantum coherence in top--antitop production at the LHC using the $l_1$-norm of coherence in the helicity basis. By combining SM predictions with the spin density matrix reconstructed from CMS measurements of the top-quark spin-correlation coefficients, we have established a direct connection between a quantum information resource and an experimentally accessible collider observable. To quantify the experimental precision of the extracted coherence, we propagated the measured uncertainties and covariance matrix of the spin-correlation coefficients using a Monte Carlo procedure.

Our results reveal a non-monotonic dependence of quantum coherence on the invariant mass and production angle. The coherence is substantial near the production threshold, is reduced in the intermediate-mass region, and becomes large again in boosted central events. The theoretical predictions show good agreement with the CMS determinations in all three kinematic regions. Importantly, the experimentally reconstructed coherence is nonzero with a significance of more than $5\sigma$ in each region. This demonstrates that the extracted coherence is statistically robust and is not merely a feature of the central values of the reconstructed spin density matrix.

A particularly important result is obtained in the intermediate-mass region, $400<M_{t\bar t}<600~\mathrm{GeV}$. Here, the CMS reconstructed coherence is $C_{l_1}=0.597\pm0.037$, while previous studies based on the same reconstructed spin density matrix found the entanglement measures, such as concurrence and negativity, to be consistent with zero~\cite{fabbrichesi2025localvsnonlocalentanglement}. Thus, the absence of experimentally certified entanglement does not imply the absence of nonzero off-diagonal interference structure in the reconstructed spin state. The sizable value of $C_{l_1}$ demonstrates that helicity-amplitude interference remains experimentally accessible in a kinematic regime where conventional entanglement measures provide limited discrimination. This establishes quantum coherence as a genuinely complementary collider observable rather than merely another quantity that can be calculated from the reconstructed density matrix.

Since the $l_1$-norm of coherence directly quantifies the off-diagonal elements of the spin density matrix, its kinematic dependence provides information on the helicity-interference structure of the underlying QCD production amplitudes. The persistence of substantial coherence in both the near-threshold and boosted-central regimes, together with its statistically significant observation across all three regions, provides a quantitative benchmark for the SM description of top-quark spin dynamics. At the same time, deviations in the coherence distribution with respect to $M_{t\bar t}$ or $\cos\Theta$ could provide complementary sensitivity to anomalous top-quark interactions, including chromomagnetic and chromoelectric dipole couplings and other higher-dimensional operators within an effective-field-theory framework.

Beyond the specific results presented here, our work demonstrates that quantum coherence can be extracted from experimentally reconstructed top-quark spin density matrices with high statistical significance and can reveal quantum interference information that is not captured by entanglement measures alone. The combination of experimentally accessible coherence, statistically controlled uncertainty propagation, and its sensitivity to the helicity-interference structure establishes quantum coherence as a promising precision observable for quantum information studies of top-quark production and for future tests of the SM and possible new physics at the LHC.}

\section*{Data availability}
The data used in the article are referenced.

\section*{Declaration of competing interest}
The authors declare that they have no known competing financial interests or personal relationships that could have appeared to influence the work reported in this paper.

\section*{Acknowledgments}
We would like to express our sincere gratitude to Prof. Mojtaba~Mohammadi~Najafabadi for his insightful and constructive comments.  S.H. and M.A. also thank the School of Particles and Accelerators at the Institute for Research in Fundamental Sciences (IPM) for their financial support of this project. H.A.Z. acknowledges the financial support provided under the postdoctoral fellowship program of Pavol Jozef Šafárik University in Košice, Slovakia, and funding by the Slovak Research and Development Agency under the contract No. APVV-24-0091.

\appendix
\section{Elements of density matrix }\label{appendixA}
In this Appendix, we present the derivation of the spin density matrix for the $t\bar{t}$ system at leading order. As discussed in the main text, the production spin density matrix in the helicity basis can be fully specified at leading order by five independent parameters: $\tilde{A}$, $\tilde{C}_{kk}$, $\tilde{C}_{nn}$, $\tilde{C}_{rr}$, and $\tilde{C}_{kr}$. The analytical expressions for these coefficients in the partonic production channels have been extensively studied and are available in the literature (see Ref.~\cite{Afik2021} for more details). For the quark–antiquark annihilation process $q\bar q$, the corresponding coefficients are given by
\begin{align}
& \tilde{A}^{q \bar{q}}=F_q\left(2-\beta^2 \sin ^2 \Theta\right), \nonumber\\
& \tilde{C}_{r r}^{q \bar{q}}=F_q\left(2-\beta^2\right) \sin ^2 \Theta, \\
& \tilde{C}_{n n}^{q \bar{q}}=-F_q \beta^2 \sin ^2 \Theta, \nonumber\\
& \tilde{C}_{k k}^{q \bar{q}}=F_q\left(2 \cos ^2 \Theta+\beta^2 \sin ^2 \Theta\right), \nonumber\\
& \tilde{C}_{r k}^{q \bar{q}}=\tilde{C}_{k r}^{q \bar{q}}=F_q \sqrt{1-\beta^2} \sin 2 \Theta,\nonumber
\end{align}
and for the gluon–gluon fusion process $gg$, we have
\begin{align}
& \tilde{A}^{g g}=F_g(\Theta)\left[1+2 \beta^2 \sin ^2 \Theta-\beta^4\left(1+\sin ^4 \Theta\right)\right], \nonumber\\
& \tilde{C}_{r r}^{g g}=-F_g(\Theta)\left[1-\beta^2\left(2-\beta^2\right)\left(1+\sin ^4 \Theta\right)\right], \nonumber\\
& \tilde{C}_{n n}^{g g}=-F_g(\Theta)\left[1-2 \beta^2+\beta^4\left(1+\sin ^4 \Theta\right)\right], \\
& \tilde{C}_{k k}^{g g}=-F_g(\Theta)\left[1-\beta^2 \frac{\sin ^2 2 \Theta}{2}-\beta^4\left(1+\sin ^4 \Theta\right)\right], \nonumber\\
& \tilde{C}_{r k}^{g g}=F_g(\Theta) \sqrt{1-\beta^2} \beta^2 \sin 2 \Theta \sin ^2 \Theta,\nonumber
\end{align}
where the normalization factors are
$F_q=1/18$, and $F_g(\Theta)=\frac{7+9 \beta^2 \cos ^2 \Theta}{192\left(1-\beta^2 \cos ^2 \Theta\right)^2}.$

\bibliography{bibliography}

\end{document}